\RequirePackage{fix-cm}
\documentclass[onecolumn,10pt]{svjour3}
\smartqed  % flush right qed marks, e.g. at end of proof
\usepackage{graphicx}
\usepackage{fixltx2e}
\usepackage[misc]{ifsym}
\usepackage{float}
\usepackage[multi-part-units=single]{siunitx}
\usepackage[font=small,labelfont=bf,tableposition=top]{caption}
\usepackage{mathtools}
\usepackage{xcolor}
\usepackage{amsmath}
\usepackage{amssymb}

\DeclareCaptionLabelFormat{andtable}{#1~#2  \&  \tablename~\thetable}
\begin{document}

\title{Spatio-Temporal Deep Learning Models for Tip Force Estimation During Needle Insertion}


\titlerunning{Spatio-Temporal Deep Learning Models for Needle Tip Forces}        % if too long for running head

\author{Nils Gessert$^1$ \and Torben Priegnitz$^1$ \and Thore Saathoff$^1$\and Sven-Thomas Antoni$^1$ \and David Meyer$^2$\and Moritz Franz Hamann$^2$\and Klaus-Peter J\"unemann$^2$ \and Christoph Otte$^1$ \and Alexander Schlaefer$^1$
}


\institute{\Letter \quad Nils Gessert, \email{nils.gessert@tuhh.de}, Tel.: +49 (0)40 42878 3389, https://orcid.org/0000-0001-6325-5092 \\ \\ $^1$ Institute of Medical Technology, Hamburg University of Technology, Hamburg, Germany \\ $^2$ Department of Urology, University Hospital Schleswig-Holstein, Kiel, Germany}

\date{Preprint. Accepted for publication in IJCARS.}

\maketitle

\begin{abstract}

\textit{Purpose} Precise placement of needles is a challenge in a number of clinical applications such as brachytherapy or biopsy. Forces acting at the needle cause tissue deformation and needle deflection which in turn may lead to misplacement or injury. Hence, a number of approaches to estimate the forces at the needle have been proposed. Yet, integrating sensors into the needle tip is challenging and a careful calibration is required to obtain good force estimates. 

\textit{Methods} We describe a fiber-optical needle tip force sensor design using a single OCT fiber for measurement. The fiber images the deformation of an epoxy layer placed below the needle tip which results in a stream of 1D depth profiles. We study different deep learning approaches to facilitate calibration between this spatio-temporal image data and the related forces. In particular, we propose a novel convGRU-CNN architecture for simultaneous spatial and temporal data processing. 

\textit{Results} The needle can be adapted to different operating ranges by changing the stiffness of the epoxy layer. Likewise, calibration can be adapted by training the deep learning models. Our novel convGRU-CNN architecture results in the lowest mean absolute error of $\SI{1.59 \pm 1.3}{\milli\newton}$ and a cross-correlation coefficient of $0.9997$, and clearly outperforms the other methods. Ex vivo experiments in human prostate tissue demonstrate the needle's application. 

\textit{Conclusions} Our OCT-based fiber-optical sensor presents a viable alternative for needle tip force estimation. The results indicate that the rich spatio-temporal information included in the stream of images showing the deformation throughout the epoxy layer can be effectively used by deep learning models. Particularly, we demonstrate that the convGRU-CNN architecture performs favorably, making it a promising approach for other spatio-temporal learning problems. 

\end{abstract}

\section{Introduction} \label{intro}

For minimally invasive procedures such as biopsy, neurosurgery or brachytherapy, needle insertion is often utilized to minimize tissue damage \cite{abolhassani.2007needle}. To facilitate accurate needle placement, needle steering, image guidance, and force estimation \cite{taylor2016medical} can be used. Accurate measurement of the forces affecting the needle tip is of particular interest, e.g., to keep track of the needle-tissue interaction and to detect potential tissue ruptures, or to generate haptic and visual feedback \cite{okamura.2004force}. Therefore, various force-sensing solutions for needles have been proposed. A simple approach is to place a force sensor externally at the needle shaft which would allow for the use of conventional force-torque sensors. However, during insertion, large frictional forces act on the needle shaft which mask the actual tip forces. Therefore, forces acting on the needle shaft either need to be decoupled from the force sensor or the force sensor needs to be placed at the needle tip \cite{kataoka.2002measurement}. This complicates building force sensors as they are usually constrained to a few millimeters in width \cite{rodrigues.2014influence} which is particularly difficult for mechatronic force sensors \cite{kataoka.2002measurement,Hatzfeld.2017}. For these reasons, fiber optical force sensors have been developed which are often smaller, biocompatible and MRI-compatible \cite{beekmans.2016fiber}. In particular, sensor concepts using Fabry-P\'erot interferometry \cite{su2011miniature,beekmans.2016fiber} or Fiber Bragg Gratings \cite{rao1997fibre,roriz2014conventional,kumar.2016detecting} have been proposed. These methods have shown promising calibration results, however, manufacturing and signal processing can be difficult when different temperature ranges and lateral forces need to be taken into account. 
Another optical method uses the imaging modality optical coherence tomography (OCT) to estimate strain and deformation from images of deformed material such as silicone \cite{kennedy2015quantitative,larin2017optical}. Also, direct force estimation from volumetric OCT data has been presented \cite{Otte.2016,gessert2018force}. Other approaches have integrated single OCT fibers that produce 1D images into needle probes \cite{kennedy2012needle}. This concept has been used to classify malignant and benign tissue \cite{Otte.2014}. \par 

In this work we present an OCT needle concept for force estimation at the needle tip. A single OCT fiber is embedded into a ferrule with an epoxy layer applied on top of it. A sharp metal tip is mounted on top of the epoxy layer to facilitate tissue insertion. Axial forces acting on the needle tip lead to a deformation of the epoxy layer which is imaged by the OCT fiber. Thus, forces can be inferred from the OCT signal. In general, this needle design is easy to manufacture and flexible as no precise fiber placement is required, the needle tip's shape can be changed and the epoxy layer's thickness and composition can be varied. Thus, softer epoxy resin could be used for application scenarios which require a high sensitivity such as microsurgery and stiffer epoxy resin could be used for large forces which occur, e.g., during biopsy \cite{beekmans.2016fiber}. However, this approach comes with challenges for calibration and force estimation. In particular, a robust, non-linear model is required which maps the deformations observed in the OCT images to forces. The signal can be understood as 2D spatio-temporal data with a spatial and a temporal dimension. Using the current observation $t_i$ and previous ones has been shown to be effective for vision-based force estimation with RGBD cameras as the current force estimate is likely reflected in prior deformation \cite{Aviles.2017towards}. In contrast to previous approaches \cite{Aviles.2017towards}, we directly learn relevant features from the data using deep learning models. This eliminates the necessity to engineer new features for other materials with different light scattering properties.

For deep learning-based spatio-temporal data processing, various method have been proposed in the natural image domain, e.g. for action recognition \cite{simonyan2014two,wang2015action} or video classification \cite{yue2015beyond,Long_2018_CVPR}. Spatial and temporal convolutions have been employed \cite{ji20133d,tran2015learning} and models using convolutional neural networks (CNNs) with a subsequent recurrent part have been proposed \cite{donahue2015long,yue2015beyond}. 
We adapt these approaches for our new 2D spatio-temporal learning problem where a series of 1D OCT A-Scans needs to be processed. In addition, we propose a novel convolutional gated recurrent unit-convolutional neural network (convGRU-CNN) architecture. 
The key idea and difference to other methods is to first learn temporal relations using recurrent layers while keeping the spatial data structure intact by using convolutions for the recurrent gating mechanism \cite{xingjian2015convolutional}. Then, a 1D CNN architecture processes the resulting spatial representation. We provide an in-depth analysis of this concept and compare it to previous approaches for deep learning-based spatio-temporal data processing. In addition, we provide qualitative results for tissue insertion experiments, showing the feasibility of the approach. 
This work extends preliminary results we presented at MICCAI 2018 \cite{gessert2018needle}. We substantially revised and extended the original paper with an extended review of the relevant literature, a more detailed explanation of our novel convGRU-CNN model and additional experiments for the model. In particular, we rerun all quantitative experiments for more consistent results, we perform additional experiments to analyze the temporal dimension and properties of the convGRU-CNN and we improve the model with recurrent batch normalization \cite{cooijmans2016recurrent} and recurrent dropout \cite{pham2014dropout}. To highlight the proposed model's advantages, we consider additional spatio-temporal deep learning models and a conventional model. We provide a more detailed comparison of the spatio-temporal models' errors and their significance. Furthermore, we provide inference times of our models to demonstrate real-time capability.

Summarized, the key contributions of this work are threefold. (1) We propose a new design for OCT-based needle tip force estimation that is flexible and easy to manufacture. (2) We present a novel convGRU-CNN architecture for spatio-temporal data processing which we use for calibration of our force sensing mechanism. (3) We show the feasibility of our approach with an insertion experiment into human ex vivo tissue.

\section{Materials and Methods}  \label{sec:methods}

\subsection{Problem Definition} \label{sec:problem}

Our force sensing needle design uses OCT which produces series of 1D images (A-scans) that need to be mapped to forces. Thus, we consider a 2D spatio-temporal learning problem with a set of $t_s$ consecutive, cropped 1D A-scans $(A_{t_{i}},A_{t_{i-1}},...,A_{t_{i-t_s}})$  with $A_{t_{i}} \in \mathbb{R}^{d_c}$ where $d_c$ denotes the crop size. The resulting matrix $M_{t_i} \in \mathbb{R}^{t_s \times d_c}$ is used to estimate a force $F_{t_{i}} \in \mathbb{R}$. 

\subsection{Deep Learning Architectures} \label{sec:model}

\begin{figure}[t]
	\centering
	\includegraphics[angle=0,width=1.0\columnwidth]{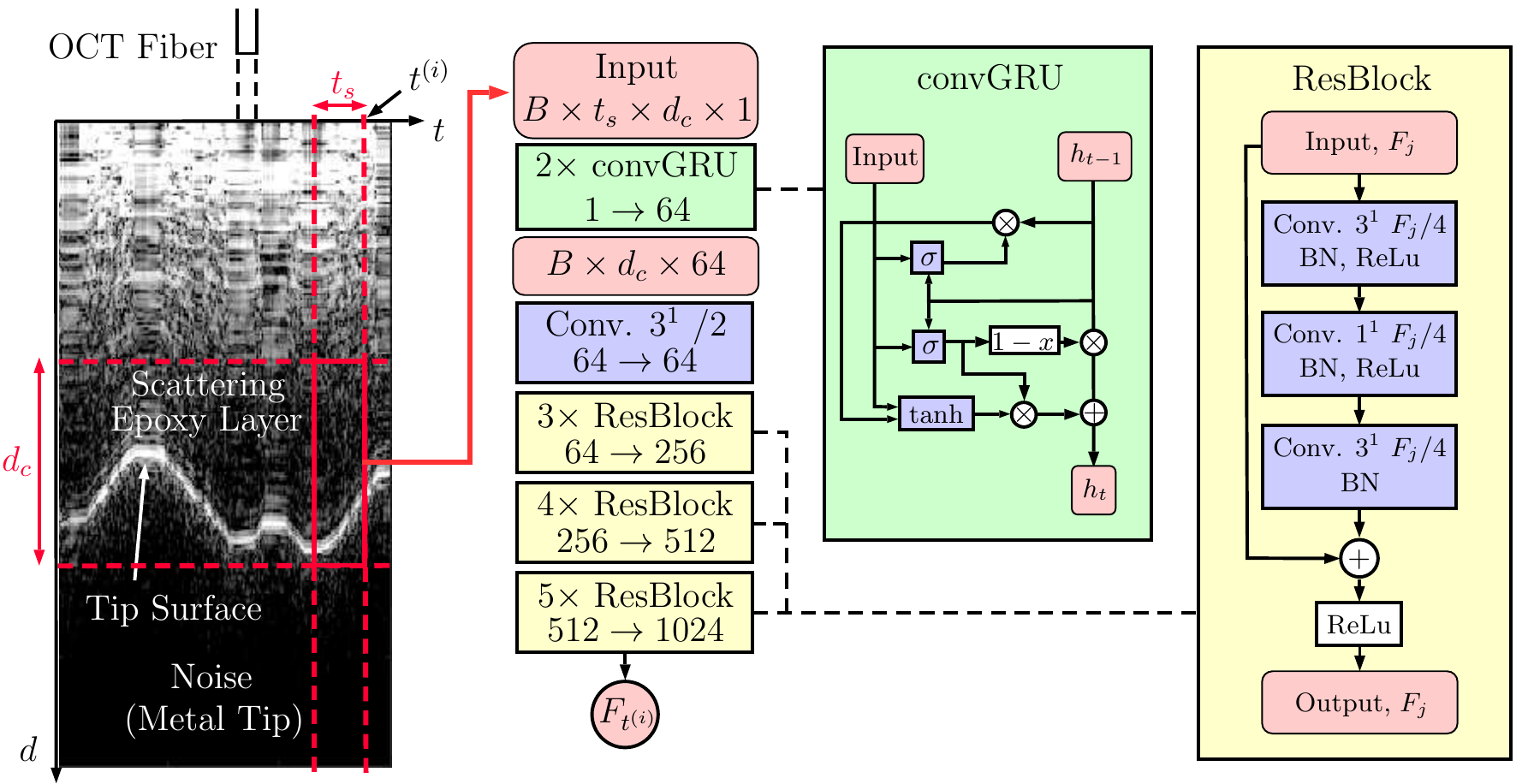}
	\caption{The convGRU-CNN model we employ.  The metal tip's flat surface at the epoxy layer cannot be penetrated by infrared light which is why that signal part is considered noise. $\sigma$ and $\tanh$ denote a convolutional gate with sigmoid and hyperbolic tangent activation function, respectively. The subsequent CNN is a ResNet-like network. The first block in a series of ResBlocks uses a stride of $2$ for the convolutions with kernel $3^1$ and increases the number of feature maps. Subsequent blocks have a stride of $1$ and keep the same feature map size. The change in the number of feature maps $F_j$ is denoted in each group of ResBlocks.} 
	\label{fig:model}
\end{figure}

We consider different deep learning models to map $M_{t_i}$ to $F_{t_{i}}$. First, we introduce our novel convGRU-CNN architecture. Then, we consider model variants that use alternative ways of data processing.

\textbf{convGRU-CNN} combines spatial and temporal data processing in a new way. First, a convolutional GRU (convGRU) takes care of temporal processing. The convGRU outputs a 1D spatial feature representation which is then processed by a ResNet-inspired \cite{He.2016} 1D CNN. The model is shown in Figure~\ref{fig:model}.
The convGRU is a combination of convLSTM and gated recurrent units \cite{cho2014learning}. We replace the matrix multiplications in the GRU with convolutions such that the output of the convGRU unit is computed as

\begin{equation}
\begin{array}{l}
z_t = \sigma (K_z * h_{t-1} + L_z * \mathit{RBN}(x_t)) \\
r_t = \sigma (K_r * h_{t-1} + L_r * \mathit{RBN}(x_t)) \\
c_t = \tanh (K_c * (r_th_{t-1}) + L_c * \mathit{RBN}(x_t)) \\
h_t = z_tc_t+(1-z_t)h_{t-1}
\end{array}
\end{equation}
where $h$ is the hidden state, $x$ is the input, $K$ and $L$ are filters, $*$ denotes a convolution, $\sigma$ denotes the sigmoid activation function and $\mathit{RBN}(.)$ denotes recurrent batch normalization \cite{cooijmans2016recurrent}. Furthermore, we employ recurrent dropout for additional regularization \cite{pham2014dropout} at the cell input with probability $p_{di} = 0.1$ and at the cell output with probability $p_{do} = 0.2$. Recurrent batch normalization and dropout are extensions to the original model presented in \cite{gessert2018needle} and the new model is named \textbf{convGRU-CNN+}. We add these augmentations to all recurrent models.
%We compare the proposed convGRU-CNN architecture to other models that are inspired by spatio-temporal deep learning approaches used in the natural image domain.

\textbf{1D CNN} processes A-Scans $A_{t_{i}}$ individually without considering a history of data which resembles a single-shot learning approach. The CNN architecture is the same ResNet-based model that is depicted in Figure~\ref{fig:model}.

\textbf{GRU} processes a set a of A-Scans, without taking spatial structure into account as it consists of three GRU layers with standard matrix multiplications being performed inside the gates. %We also use recurrent batch normalization and recurrent dropout for this model.

\textbf{CNN-GRU} also uses a set a of A-Scans and follows the classic approach of first performing spatial processing and feature extraction with a CNN and then temporal processing with a recurrent model. The CNN part is the ResNet-based model as shown in Figure~\ref{fig:model} and the recurrent part is a two-layer GRU. 

\textbf{2DCNN} is fed with a set a of A-Scans and performs data processing with convolutions over both the spatial and temporal dimension. This architecture also follows the ResNet-like CNN part shown in Figure~\ref{fig:model}. The kernels are of size $3 \times 3$ and strides are used to simultaneously reduce the spatial and the temporal dimension. 

\textbf{GRU-CNN} is a variant of convGRU-CNN with normal GRU cells. Here, the A-Scans are directly treated as feature vectors. This architecture is used to demonstrate the necessity of using convolutional GRUs when performing temporal processing first.

\textbf{CNN-convGRU} is a variation of CNN-GRU with convGRU cells. Before the global average pooling operation, the convGRU cells perform temporal processing while keeping the spatial structure that resulted from CNN processing. Afterwards, global average pooling is applied and the resulting feature vector is fed into the output layer. This architecture serves as a comparison to convGRU-CNN in terms of the position of the convGRU units in the network.

\textbf{MIP-GPM} is a simple reference model using classic feature extraction with a Gaussian process regression model \cite{rasmussen2004gaussian}. We extract the needle tip's high-intensity surface using 1D maximum intensity projection (MIP) on the median-filtered A-scans. The normalized pixel index of the MIP represents a simple feature that captures deformation. This scalar feature serves as a comparison to the deep learning models.

All networks are trained end-to-end. We use the Adam algorithm for optimization with a batch size of $B=100$. Our implementation uses Tensorflow \cite{Abadi.2016}. The initial learning rate is $l_r = \num{e-4}$. We halve the learning rate every $\num{30}$ epochs and stop training after $\num{300}$ epochs. 

\begin{figure}[t]
	\centering
	\includegraphics[width=1.0\columnwidth]{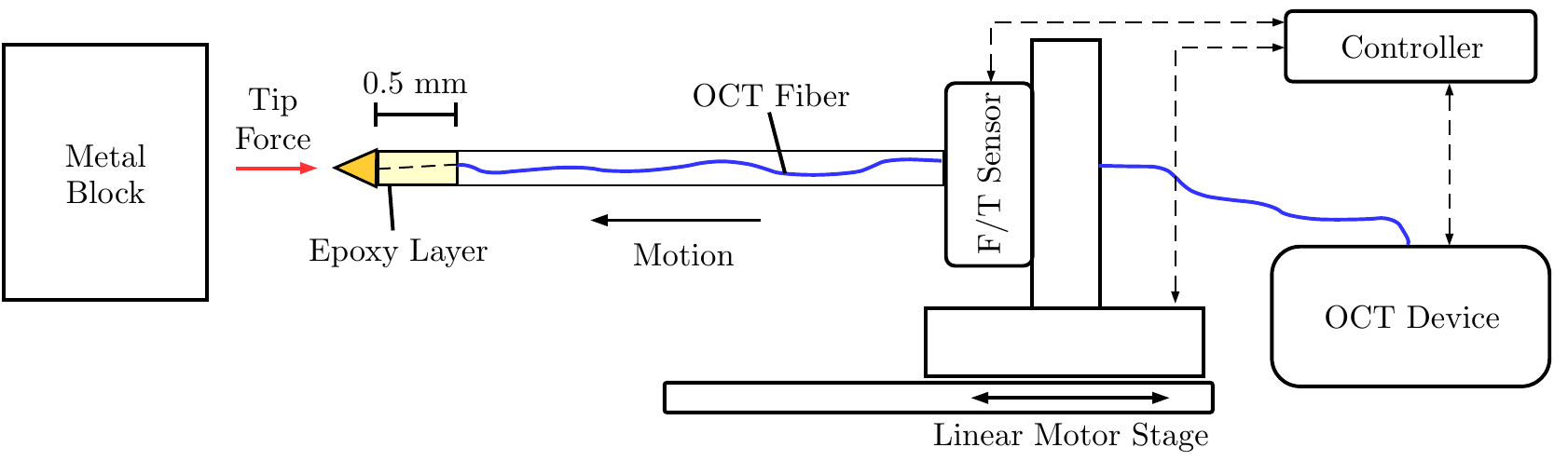}
	\caption{Schematic drawing of the needle and the calibration setup. Not to scale. The needle contains an OCT fiber that images a deformable epoxy layer below the needle tip. Forces are measured by the force sensor at the base. The setup is moved with a linear stage.}
	\label{fig:setup}
\end{figure}

\subsection{Needle Design and Experimental Setup}

\begin{figure}[t]
	\centering
	\includegraphics[width=0.199\columnwidth]{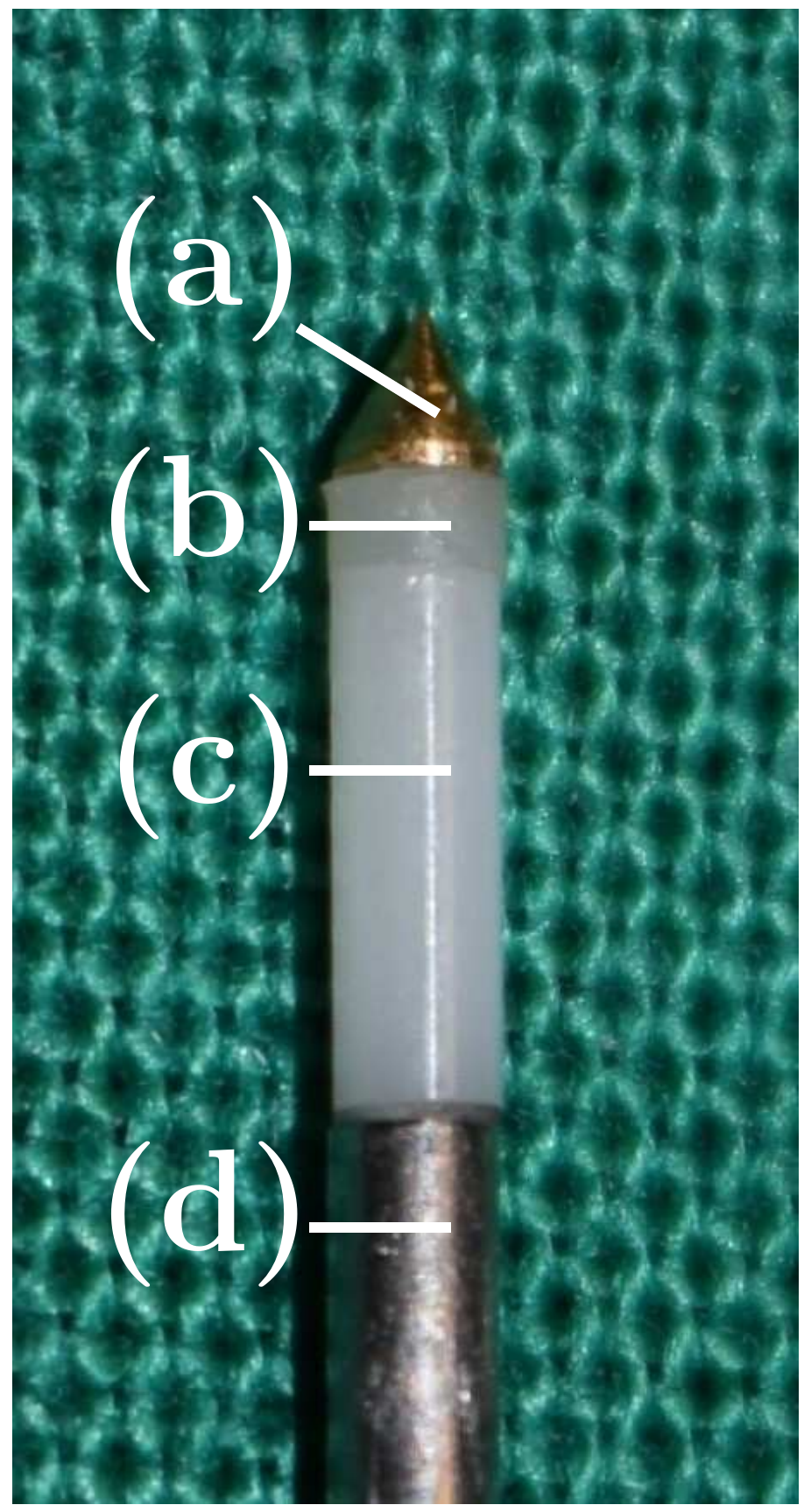}
	\includegraphics[width=0.792\columnwidth]{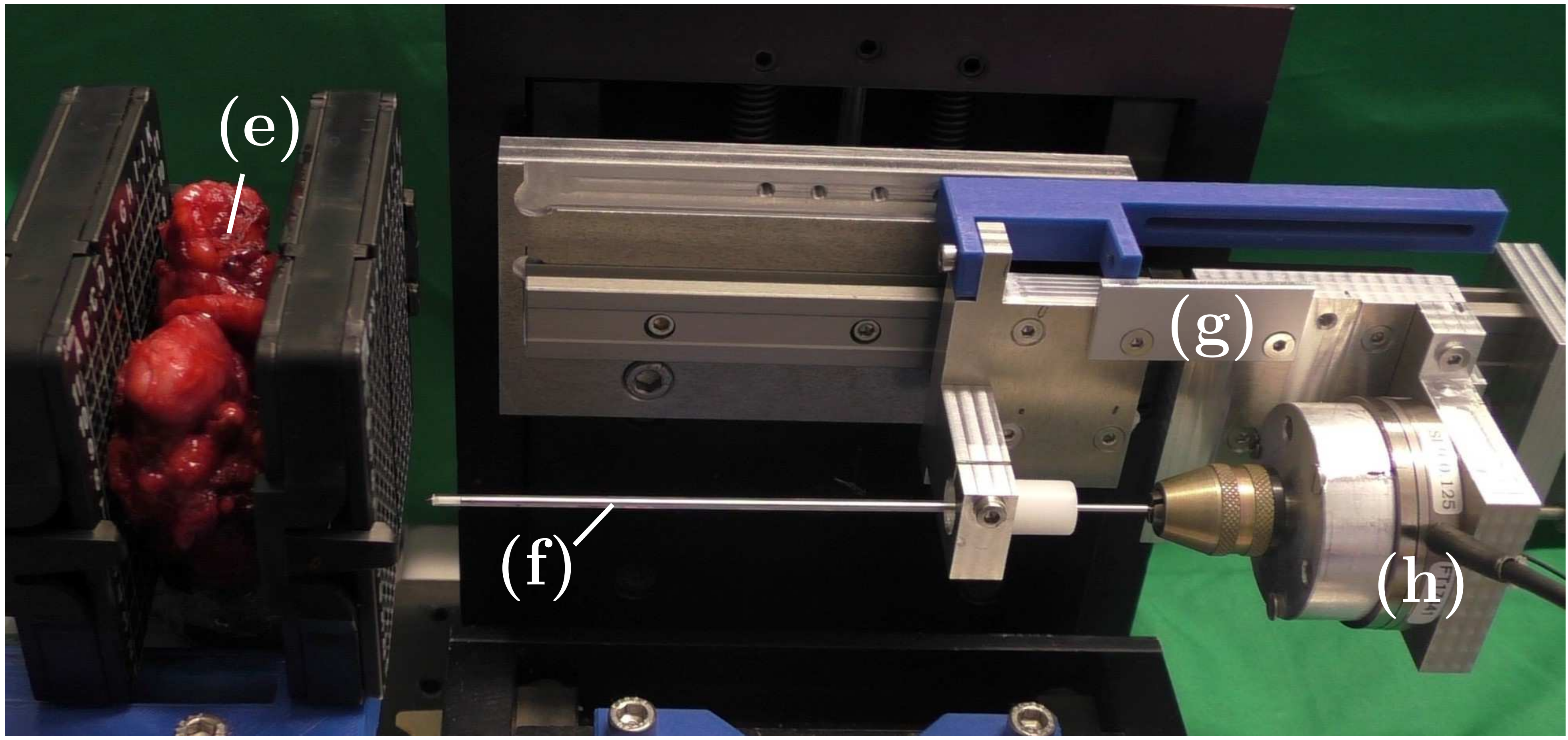}
	\caption{Needle design (left) and photograph of the experimental setup for the prostate insertion experiment (right). The brass tip (a) is attached to the epoxy layer (b) which is glued to the ferrule with the embedded OCT fiber (c). The ferrule is attached to the needle base (d) with a diameter of $\SI{1.25}{\milli\metre}$. For the prostate (e) insertion experiment, the needle is decoupled with a shielding glass tube (f). The linear stage (g) moves the needle and the force sensor (h) acquires reference data.}
	\label{fig:setup_pic}
\end{figure}

Our proposed needle tip force sensing mechanism and calibration setup are shown in Figure~\ref{fig:setup}. The needle's base is a ferrule with a diameter of \SI{1.25}{\milli\metre} which holds the OCT fiber. 
On top, we apply an epoxy resin layer with a height of \SI{0.5}{\milli\metre} using Norland Optical Adhesive (NOA) 63. On top of the layer, a cone-shaped brass tip is attached. The epoxy resin layer's stiffness is varied by mixing the resin with different concentrations of NOA 1625.
The needle's OCT fiber is attached to a frequency domain OCT device (Thorlabs Telesto I). A force sensor (ATI Nano43) for ground-truth annotation is mounted between the needle and a linear stage that moves the needle along its axial direction. 
For calibration, the tip is deformed with random magnitude and velocity to create a large dataset with extensive force variations being covered.
Next, we validated the needle in tissue insertion experiments, see Figure~\ref{fig:setup_pic}. Obtaining ground-truth tip forces is challenging for this case as the force sensor at the base measures both axial tip forces and friction forces acting on the shaft. Therefore, we use a shielding tube which is decoupled from the needle and the force sensor. This allows for measurement of axial tip forces for comparison to our needle tip sensing mechanism. Note, that the shielding tube is a workaround for validation experiments but not for practical application as the stiff tube would increase trauma. 
We perform insertion experiments into a freshly resected human prostate. In the supplementary material, video, ultrasound, force and OCT signal recordings are provided.

\subsection{Data Acquisition and Datasets}

The OCT device we use is a frequency-domain-OCT which uses interferometry with near infrared light to acquire 1D depth profiles (A-Scans) with a rate of $\SI{5500}{\hertz}$. The light's wavelength of $\SI{1325}{\nano\metre}$ allows for imaging of the inner structure of scattering materials with up to $\SI{1}{\milli\metre}$ depth. 
The force sensor for ground-truth annotation acquires data at $\SI{500}{\hertz}$. Therefore, the OCT and force sensor data streams need to be synchronized and matched. We use the streams' timestamps for synchronization and nearest neighbor interpolation to assign an A-scan to each force measurement. To construct a sequence, we add $t_s$ previous A-scans to each A-scan with an assigned force value. 

We acquire calibration datasets for three needles with different stiffness of the epoxy layer. The datasets each contain approximately $90000$ sequences of A-scans, each labeled with a scalar, axial force. We use $\SI{80}{\percent}$ of the data for training and validation and $\SI{20}{\percent}$ for testing. There is no overlap between the sequences from the different sets.
We tune hyperparameters based on validation performance. 
In terms of metrics we report the mean absolute error (MAE) in $\si{\milli\newton}$ with standard deviation, the relative MAE (rMAE) with standard deviation and correlation coefficient (CC) between predictions and targets. To ensure consistency, we repeat all experiments five times and provide the mean values over all runs. We test for significant difference in the median of the models' absolute errors with the Wilcoxon signed-rank test ($\alpha = \SI{5}{\percent}$ significance level). Furthermore, we provide the inference time (IT) in $\si{\milli\second}$ of each model for a single forward pass, averaged over $100$ repetitions.

\section{Results}

First, we report the results for the three different needles with different stiffness of the epoxy layer. Stiffness increase from needle one to three. The results with the corresponding maximum force magnitudes are shown in Table~\ref{tab:res_calibs}. With increasing stiffness, the MAE increases, as the overall covered force range increases. Between needle 1 and 2, the rMAE increases by a factor of $1.33$. Between needle 2 and 3, the rMAE increases by a factor of $1.29$. The CC remains similar among the needles. 

\begin{table}
	\centering
	{\setlength{\tabcolsep}{1.5em}
	\begin{tabular}{l l l l l}
	 & MAE & rMAE & CC & Max \\ \hline
	Needle 1 & $1.59 \pm 1.3$ & $0.0199 \pm 0.0172$ & $0.9997$ & $379$ \\
	Needle 2 & $7.12 \pm 5.8$ & $0.0266 \pm 0.0226$ & $0.9995$ & $974$ \\	
	Needle 3 & $22.97 \pm 20.6$ & $0.0345 \pm 0.0301$ & $0.9991$ & $3202$ \\
 \hline \\
	\end{tabular}}
	\caption{Comparison of needles with different epoxy layer stiffnesses. The convGRU-CNN+ model was used for this experiment.}
	\label{tab:res_calibs}
\end{table}

\begin{figure}[t]
	\centering
	\includegraphics[width=0.8\columnwidth]{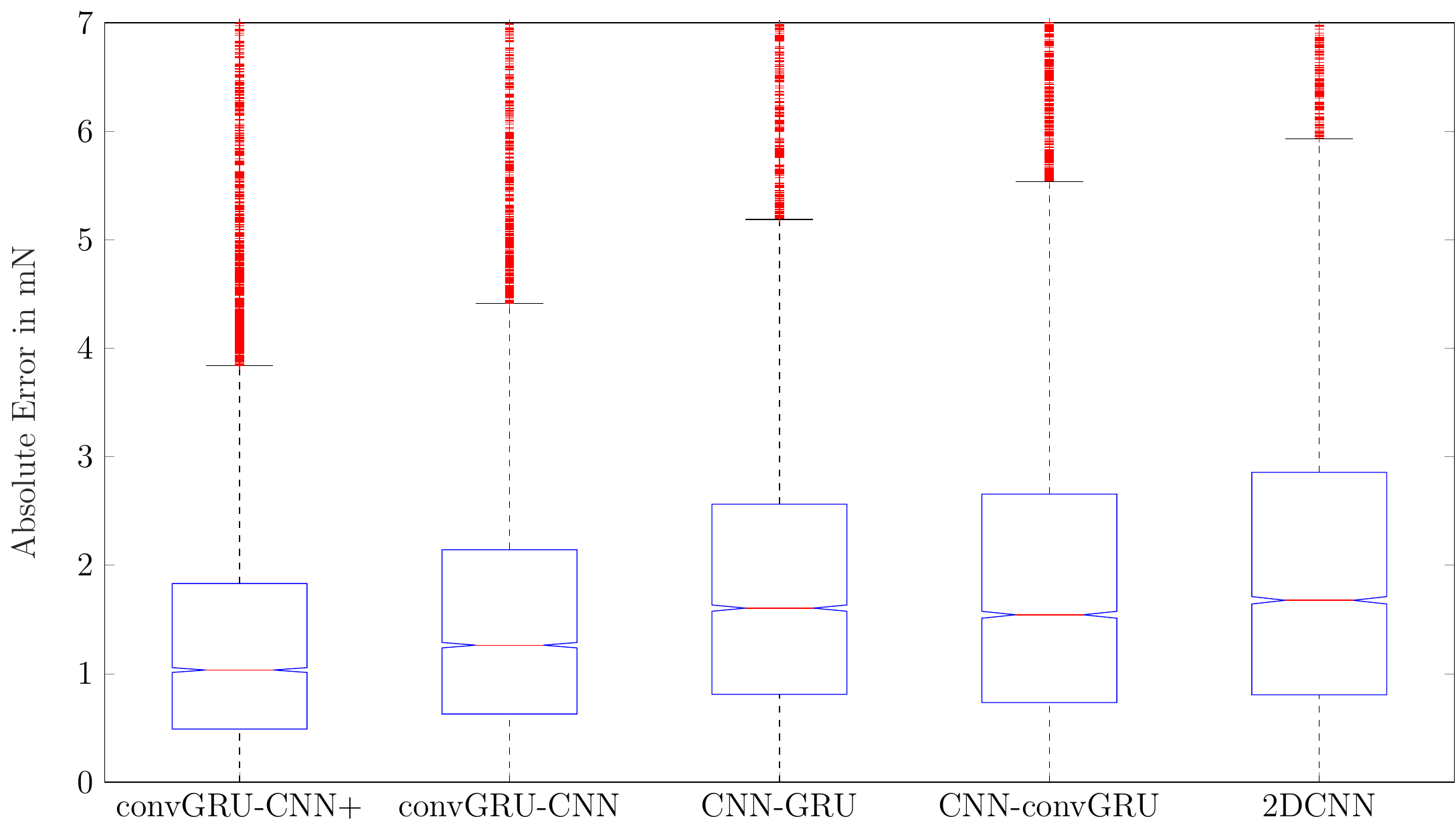}
	\caption{Boxplots of the absolute errors for the top-performing spatio-temporal deep learning models. The red line marks the median, the boxes' bottom and top line mark the 25th and 75th percentiles, respectively. Red marks above the whiskers represent outliers. Notches around the median mark comparison intervals where non-overlapping intervals between boxplots indicate different medians at $\SI{5}{\percent}$ significance level. With respect to the frequency of outliers, consider that the errors are likely not normally distributed.}
	\label{fig:boxplots}
\end{figure}

Next, we compare our proposed convGRU-CNN model to other spatio-temporal deep learning methods. The results are shown in Table~\ref{tab:models}. Overall, the models that perform spatio-temporal processing (convGRU-CNN+, convGRU-CNN, CNN-GRU, CNN-convGRU, 2DCNN) clearly outperform  1DCNN and GRU. Overall, convGRU-CNN+ performs best. Boxplots in Figure~\ref{fig:boxplots} show a more detailed analysis of the spatio-temporal deep learning models. The null hypothesis of an equal median for the absolute errors of convGRU-CNN+ can be rejected with $p$-values of $\num{6.15e-10}$, $\num{1.32e-11}$, $\num{7.72e-12}$ and $\num{1.16e-12}$ compared to convGRU-CNN, CNN-GRU, CNN-convGRU and 2DCNN, respectively.

In terms of inference time, the spatio-temporal deep learning models can provide predictions with approximately $\SI{100}{\hertz}$. The fastest spatio-temporal deep learning model is 2DCNN with an IT of $\SI{8.6}{\milli\second}$ and the overall fastest model is GRU with an IT of $\SI{2.5}{\milli\second}$. Note that these values are highly hardware (NVIDIA GTX 1080 Ti) and software (Tensorflow) dependent.

\begin{table}
	\centering
	{\setlength{\tabcolsep}{1.0em}
	\begin{tabular}{l l l l l}
	 & MAE & rMAE & CC & IT \\ \hline
	\textbf{convGRU-CNN+} & \boldmath $1.59 \pm 1.3$ & \boldmath $0.0199 \pm 0.0172$ & \boldmath $0.9997$ & $10.3 \pm 1.5$ \\
	convGRU-CNN & $1.75 \pm 1.5$ & $0.0211 \pm 0.0182$ & $0.9996$ & $9.6 \pm 1.7$ \\
	GRU & $3.02 \pm 3.7$ & $ 0.0377 \pm 0.0479$ & $0.9982$ & $\boldmath 2.5 \pm 0.4$ \\	
	1DCNN & $3.26 \pm 3.9$ & $0.0393 \pm 0.0482$ & $0.9980$ & $6.9 \pm 1.3$ \\ 		
	CNN-GRU & $2.01 \pm 3.2$ & $0.0247 \pm 0.0411$ & $0.9989$ & $9.5 \pm 1.4$ \\
	CNN-convGRU & $2.03 \pm 3.3$ & $0.0242 \pm 0.0426$ & $0.9990$ & $11.1 \pm 1.7$ \\
	2DCNN & $2.11 \pm 3.5$ & $0.0255 \pm 0.0438$ & $0.9987$ & $8.6 \pm 1.5$ \\
	GRU-CNN & $11.79 \pm 8.6$ & $0.1253 \pm 0.1001$ & $0.9948$ & $10.1 \pm 1.4$ \\ 
	MIP-GPM & $45.38 \pm 38.7$ & $0.4820 \pm 0.4114$ & $0.7767$ & $14.8 \pm 2.2$ \\ \hline \\	
	\end{tabular}}
	\caption{Comparison of several architectures. Needle 1 was used for this experiment.}
	\label{tab:models}
\end{table}

\begin{figure}[t]
	\centering
	\includegraphics[width=1.0\columnwidth]{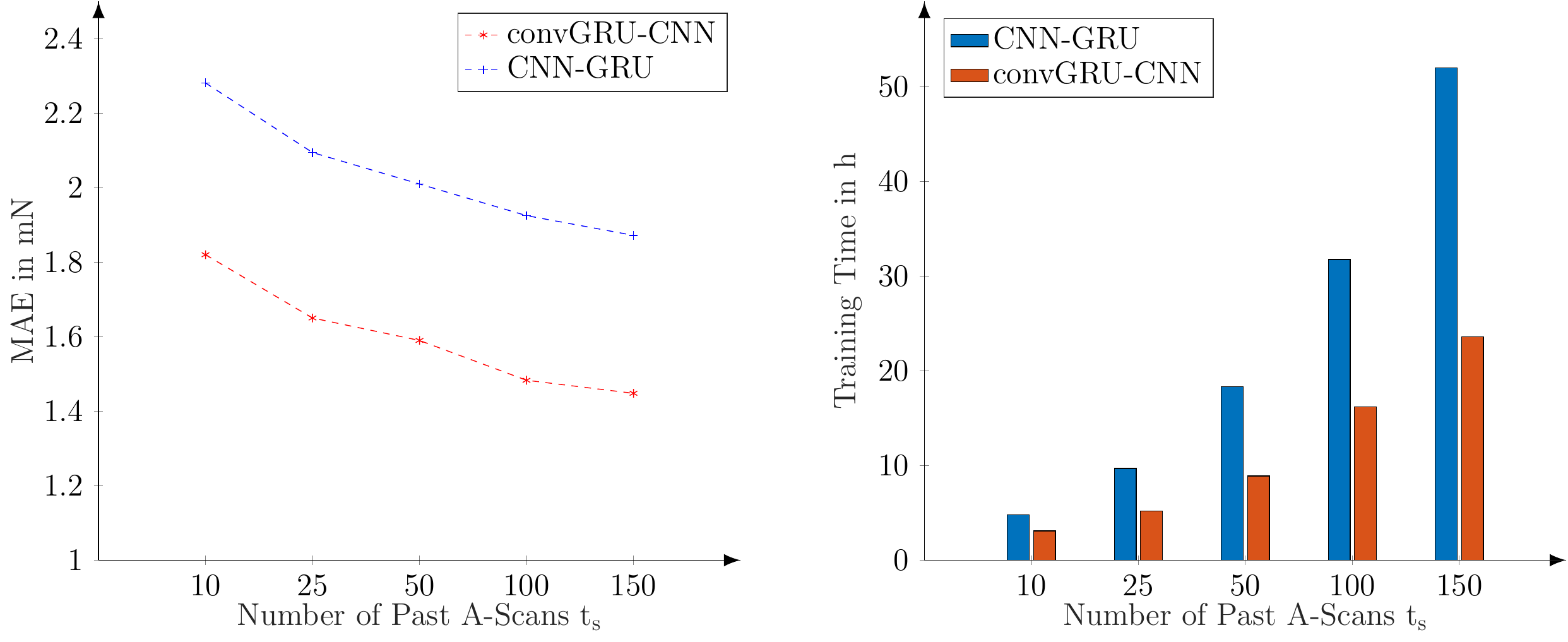}
	\caption{Comparison between convGRU-CNN+ and CNN-GRU for different numbers of timesteps $t_s$. The calibration data for needle 1 was used for this experiment.}
	\label{fig:temp}
\end{figure}

The previous results showed a clear performance increase for joint spatio-temporal processing. Therefore, we perform experiments to analyze the effect of the temporal dimension. In Figure~\ref{fig:temp} we show results for different $t_s$ and the associated training durations with our convGRU-CNN+ and CNN-GRU model. Increasing $t_s$ leads to improved performance with a lower MAE for both models. With increasing $t_s$, the overall training time also increases substantially. Across all values for $t_s$, the training time of convGRU-CNN+ is lower than the time for CNN-GRU. 

\begin{figure}[t]
	\centering
	\includegraphics[width=1.0\columnwidth]{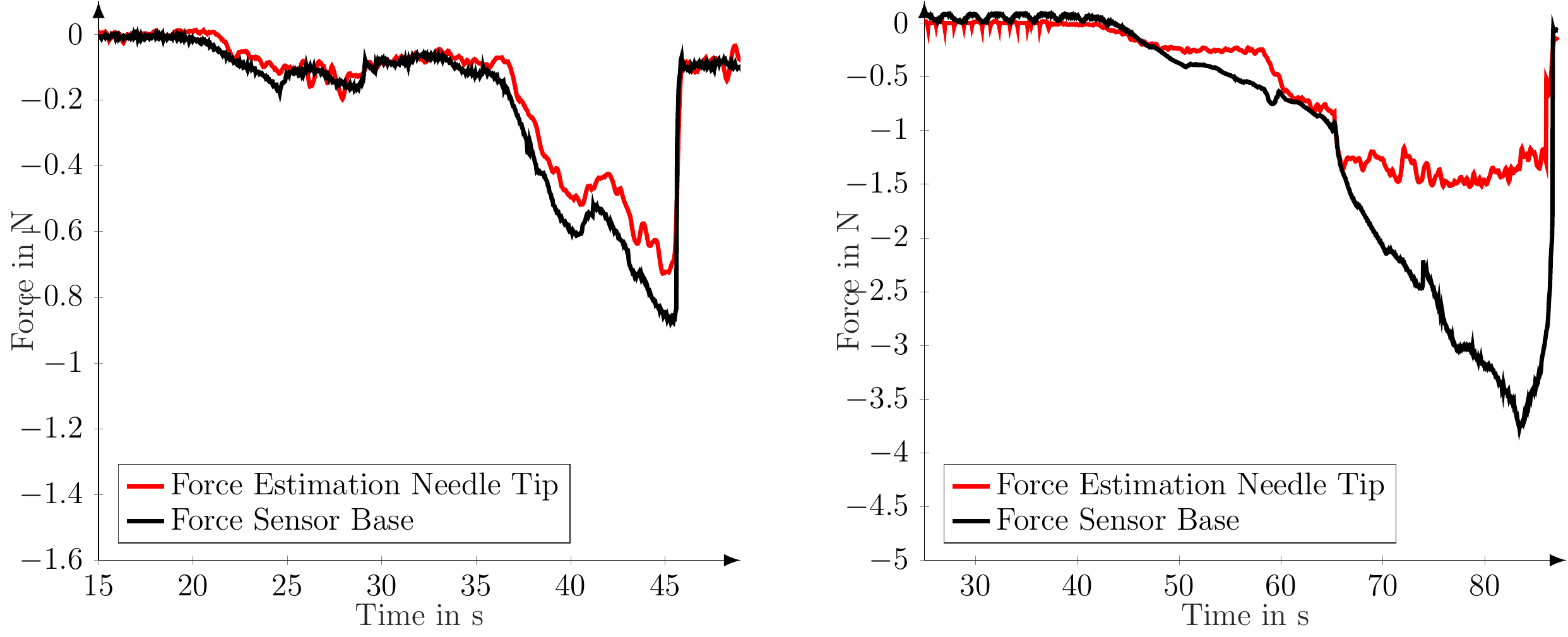}
	\caption{Predicted and measured force values are shown for an insertion with the shielding tube (left) and without (right). For the case without tube, differences between needle tip force estimation and force sensor is caused by friction. Needle 2 and convGRU-CNN+ were used for this experiment.}
	\label{fig:tissue}
\end{figure}

Last, we present results for the needle insertion experiments shown in Figure~\ref{fig:tissue}. We performed one experiment with the shielding tube and without. When using the decoupled tube, the force sensor's measurements for ground-truth annotation closely match the the values predicted by our model. Without the tube, friction forces induce a large difference between measurements and predictions. 

\section{Discussion}

We present a new technique for needle tip force estimation using an OCT fiber embedded into a needle that images the deformation of an epoxy layer. OCT has been used for multiple needle-based tissue classification scenarios \cite{Otte.2014,muller2015prostate} which could lead to more wide-spread application in clinical settings. 
Our needle is flexible in design and easy to manufacture. This is highlighted by our results for three needles with epoxy layers of different stiffness. With increasing stiffness, the rMAE increases slightly by $\SI{30}{\percent}$ between needles which indicates that there is a decrease in relative performance for stiffer needles but the decrease appears to be bounded as it is similar for needle 1 and 2 and needle 2 and 3. Also, the CC remains high in a range of $0.9997$ to $0.9991$. This indicates that our method generalizes well for different epoxy stiffness levels. Overall, this allows for flexible adaptation of our needle to scenarios with different requirements for force sensitivity and range.  

The OCT fiber within the needle produces series of A-Scans that can be treated as spatio-temporal data, i.e. 1D images over time. To process this type of data we propose a novel convGRU-CNN+ architecture. The model performs both temporal and spatial processing and outperforms the pure temporal GRU and pure spatial 1D CNN with an MAE of $1.59 \pm 1.3$ compared to an MAE of $3.02 \pm 3.7$ and $3.26 \pm 3.9$, respectively. Also, we compared to the spatio-temporal models CNN-GRU, CNN-convGRU and 2DCNN which are variants adopted from the natural image domain \cite{donahue.2015long,ballas2015delving,xingjian2015convolutional}. The three models are closer in terms of performance but overall, convGRU-CNN+ performs best. Notably, the differences in the median of the errors are significant which is also highlighted by the boxplots showing the test set error distribution in Figure~\ref{fig:boxplots}. 

The key difference between all spatio-temporal deep learning models is that convGRU-CNN(+) and GRU-CNN first perform temporal processing, then spatial processing, CNN-GRU and CNN-convGRU first performs spatial, then temporal processing and 2DCNN performs concurrent processing. Overall, our proposed model significantly outperforms all other variants. The lower performance of the previous spatio-temporal models CNN-GRU \cite{donahue.2015long} and CNN-convGRU \cite{ballas2015delving} indicates that temporal processing followed by spatial processing is preferable for the problem at hand. To highlight the necessity of convGRU units, we consider GRU-CNN without convolutional gates. The MAE is significantly higher which demonstrates the necessity to preserve the spatial structure during temporal processing. In addition, we show that recurrent dropout and recurrent batch normalization can improve the spatio-temporal models' performance further. For reference, MIP-GPM shows that conventional feature extraction without extensive engineering cannot match deep learning models' performance for this problem.

Furthermore, we perform a more detailed analysis of our convGRU-CNN+ model compared to the more common CNN-GRU model. The results in Figure~\ref{fig:temp} show a decrease of the MAE when a longer history of A-Scans is considered. This highlights the value of exploiting temporal information for force estimation. However, this improvement is bought with a substantial increase in training time as the computational effort increases. For example, for convGRU-CNN+, using $t_s = 100$ instead of $t_s = 50$ previous measurements leads to a performance increase of $\SI{7}{\percent}$ and an increase in training time of $\SI{82}{\percent}$. Training time is an important aspect to consider for application as newly designed needles will require an initial calibration, i.e. model training. If a new needle with adjusted epoxy layer for a particular force range needs to be available quickly, performance needs to be traded off against shorter training times. 
Overall, both models benefit similarly from the additional temporal information, however, convGRU-CNN+ trains faster. This is due to the convGRU units which have significantly fewer parameters than their GRU counterpart in CNN-GRU.

Besides performance and training time, the models' inference time is important for application and real-time feedback of forces. Overall, the high-performing spatio-temporal deep learnings models can process samples at $\SI{100}{\hertz}$ which indicates real-time capability. Notably, 2DCNN is almost as fast as 1DCNN due to the fact that 2D convolutions are much more common and highly optimized in Tensorflow and CUDA. Thus, our convGRU-CNN model's inference time could improve further with software optimization as a 1D CNN is also part of the model. Also, note that inference times are hardly affected by the number of previous measurements $t_s$. After initial processing, the recurrent part of the models stores previous information in its cells' states and only requires one additional sample to be processed at each step.

Also, it is important how the estimated forces are transferred to the physician effectively \cite{kitagawa2005effect}. Previous approaches used a haptic feedback device \cite{meli2017experimental} or visual feedback \cite{aviles2018sensory} to provide the forces to the physicians. In this context, the application scenario and required force resolution are important as rough estimates might be sufficient for qualitative feedback methods. Other settings such as retinal microsurgery require highly accucrate force measurements \cite{he2014submillimetric} where our high-performing models might be particularly beneficial. Thus, future work could examine our proposed force estimation method in the context of different application scenarios and force feedback methods.

Last, we validated our needle design in a realistic insertion experiment with a freshly resected human prostate. Our results in Figure~\ref{fig:tissue} show that the needle tip forces closely match the actual, decoupled, base measurements. 
While the decoupling is not perfect, we can show that our method accurately captures events such as ruptures. Without tip measurements or decoupling, large friction forces overshadow the actual tip forces. Overall, the experiments show that our method is usable for actual force estimation in soft tissue.

\section{Conclusion}

We introduce a new method to measure forces at a needle tip. Our approach uses an OCT fiber imaging the deformation of an epoxy layer to infer the force that acts on the needle tip. 
To map the OCT data to forces, we propose a novel convGRU-CNN+ architecture for spatio-temporal data processing. We provide an analysis of the model's properties and we show that it outperforms other deep learning methods. Furthermore, validation experiments in ex vivo human prostate tissue underline the method's potential for practical application.
For future work, our convGRU-CNN+ architecture could be employed for other spatio-temporal learning problems. 

\section*{Compliance with Ethical Standards}

\small \textbf{Funding:} This work was partially supported by DFG grants SCHL 1844/2-1 and SCHL 1844/2-2.

\small \textbf{Conflict of Interest:} %\section*{Conflict of Interest}
The authors declare that they have no conflict of interest.

\small \textbf{Ethical Approval:} %\section*{Ethical Approval}
All procedures performed in studies involving human participants were in accordance with the ethical standards of the institutional and/or national research committee and with the 1964 Helsinki declaration and its later amendments or comparable ethical standards.

\small \textbf{Informed Consent:} %\section*{Informed Consent}
Informed consent was obtained from all individual participants included in the study.

%\begin{acknowledgements}
%\end{acknowledgements}

\bibliographystyle{spmpsci} 
\bibliography{egbib}   % name your BibTeX data base

\end{document}